\documentclass[]{elsart}
\usepackage{psfig}

\newcommand{\SeBa}{\mbox{{\tt SeBa}}}

\newcommand{\rsun}{\mbox{${\rm R}_\odot$}}
\newcommand{\msun}{\mbox{${\rm M}_\odot$}}
\newcommand{\mmean}{\mbox{${\langle m \rangle}$}}

\def\apgt{\ {\raise-.5ex\hbox{$\buildrel>\over\sim$}}\ }
\def\aplt{\ {\raise-.5ex\hbox{$\buildrel<\over\sim$}}\ }

\begin{document}

\runauthor{Simon Portegies Zwart}
\begin{frontmatter}
\title{McScatter: a Simple Three-Body Scattering Package with Stellar
Evolution}
\author[Edinburgh]{Douglas C. Heggie}
\author[Amsterdam]{Simon Portegies Zwart}
\author[Monash]{Jarrod R. Hurley}

\address[Edinburgh]{The University of Edinburgh, James Clerk Maxwell Building,
Mayfield Road, Edinburgh, EH9 3JZ, UK}

\address[Amsterdam]{Astronomical Institute 'Anton Pannekoek' and
Section Computational Science, University of Amsterdam, Amsterdam,
the Netherlands}

\address[Monash]{Centre for Stellar and Planetary Astrophysics, Monash
University, Victoria, 3800 Australia}


\begin{abstract}
We describe a simple computer package which illustrates a method of
combining stellar dynamics with stellar evolution.  Though the method
is intended for elaborate applications (especially the dynamical
evolution of rich star clusters) it is illustrated here in the context
of three-body scattering, i.e. interactions between a binary star and
a field of single stars.  We describe the interface between the
dynamics and the two independent packages which describe the internal
evolution of single stars and binaries.  We also give an example
application, and introduce a stand alone utility for the visual
presentation of simulation results.  
\end{abstract}
\begin{keyword}
Stellar dynamics ---
          Methods: numerical ---
          Techniques: miscellaneous  ---
          binaries: close ---
          Stars: evolution 
\end{keyword}
\end{frontmatter}

\section{Introduction}

Many methods exist for modelling the dynamical evolution of rich star
clusters (see Heggie \& Hut 2003, ch.9), but {\sl very} few include
anything other than the crudest model of stellar evolution.  Indeed
realistic evolution of single and binary stars is confined to the
$N$-body codes developed by Aarseth and colleagues (Aarseth 2004) and
those included in the {\tt starlab} package (Portegies Zwart et al
2001, Appendix B).  For the efficient modelling of globular star
clusters, faster methods are desirable, but none yet incorporates
stellar and binary evolution of any sophistication (see however
Portegies Zwart et al.\,1997; Ivanova et al.\, 2004).  Our aim in this
paper is to facilitate this development, by showing how the stellar
evolution modules in $N$-body codes can be incorporated into other
codes.  For this purpose we have developed a {\sl very} simple
package, called McScatter,
which simulates the evolution of a single binary in dynamical
interactions with a field of single stars.  Exchanges are allowed, and
therefore all essential aspects of dynamics which affect binaries in
star cluster simulations are incorporated.  We begin with a
description of the dynamics, and then proceed to those aspects of the
stellar and binary evolution modules which have to be understood for
our purposes.  The code we are describing is publicly available at
{\tt http://manybody.org/manybody/McScatter.html} \footnote{The {\tt
SeBa} version of McScatter requires the {\tt starlab} package to be installed on your system, which is available via {\tt
http://manybody.org/manybody/starlab.html}}.

\section{Description of McScatter}

The main purpose of McScatter is to illustrate how stellar evolution
is to be interfaced with stellar dynamics. We describe the dynamical part
in outline only (in the following subsection).  Subsequent subsections
describe the interface with the stellar evolution packages in somewhat
greater detail.

The essential structure of the code is a loop (Table 1) in which it is
decided whether or not a significant scattering event occurs within a
timestep. The timestep size is selected based on the encounter rate
and the evolutionary state of the binary.  If no scattering event
occurs during the selected timestep the binary is evolved to the
current time and the new timestep is determined.  On the other hand,
if a scattering event occurs with a single star of some mass selected
from the initial mass function, this star is then evolved to the time
of the encounter, as explained in \S\,\ref{Sect:dynamics}.

Note that, if a
binary is affected (e.g. by exchange) in any encounter, the updating
of the binary takes place during the following pass through the main
loop.  This is done only in order to keep the updating of the binary
parameters in one place in the code.

\vskip0.5truein
\centerline{Table 1: Flow-Chart}
\begin{verbatim}
Initialize binary with primary mass, secondary mass, 
                       semi-major axis and eccentricity.
compute maximum encounter rate 

loop for a Hubble time
BEGIN
  determine time step (dt)
  check whether encounter occurs in this timestep with third body of
	ZAMS mass (m): set flag SCATTER and encounter time
  update binary parameters for previous encounter (if flagged)
  evolve binary to time (t+dt) or time of encounter

  if(SCATTER == TRUE)
  BEGIN
    evolve encountering star to moment of encounter
          (note that the mass that was set was the ZAMS mass; it  
           may be affected by stellar evolution.)
    perform scatter between binary and third star 
    print intermediate result
  END
  print final result
END
\end{verbatim}

\subsection{Dynamics}\label{Sect:dynamics}

We adopt a very simple scattering cross section $\Sigma$, based on the
geometrical cross section of the binary and gravitational focusing,
and normalised to give the correct result in the well studied case of
equal masses, i.e.
\begin{equation}
\Sigma = \frac{5\pi}{16}\sqrt{\frac{\pi}{3}}\frac{AGM_{123}a}{V^2},
\end{equation}
where $A = 21$ (Spitzer 1987, Sec.6.1b), $G$ is the constant of
gravitation, $M_{123}$ is the sum of the masses of the stars
participating in the 3-body encounter, $a$ is the semi-major axis of
the binary, and $V$ is the relative speed (at infinity) of the third
star and the centre of mass of the binary.

For the single stars we assume a power law initial mass function 
\[
f(m) dm \propto m^{-\alpha} dm
\] 
between some minimum $m_-$ and upper limit $m_+$, and a space number
density $n$.  We suppose that $V$ has a Maxwellian distribution in
equipartition, in the sense that, if $\sigma$ is the one-dimensional
velocity dispersion of stars of average mass $\mmean$, the mean square
value of $V$ is
\[
\langle V^2\rangle = \frac{3 \mmean M_{123}}{M_{12}m_3}\sigma^2,
\]
where $M_{12}, m_3$ are the total mass of the binary and the mass of
the single star, respectively.

From these formulae the code estimates an upper bound on the rate of
encounters, and we choose the time step $\delta t$ so that the corresponding
average number of encounters is less than $0.1$.  The following simple rejection
method then determines simultaneously, for the current time step, whether
an encounter occurs, and the value of $m_3$.  The procedure uses the
fact that the probability of
occurrence of an encounter is 
\begin{equation}
  \label{eq:P}
P = \delta t\int_{m_-}^{m_+}\langle\Sigma V\rangle nf(m_3) dm_3,
\end{equation}
where the average is taken over the distribution of $V$.
Two random numbers $X,Y$ are chosen independently from uniform distributions on the
ranges $(m_-,m_+)$ and $(0,1)$, respectively, and an encounter with
$m_3 = X$ is deemed to occur if $Y$ is smaller than the integrand in eq.(\ref{eq:P})

If an encounter occurs, the time of the encounter is chosen randomly
within the time step, and the evolution of the binary and the
encountering star are updated.  (An error is committed here, because
the evolution will change the assumed mass of the encountering star
and possibly the orbital parameters of the binary.)  The manner in
which the stellar evolution is implemented is discussed in subsequent
sections.  Here we mention the remaining steps of the dynamical
evolution.

The adopted probability of an exchange is based on results of Heggie,
Hut \& McMillan (1996).  For simplicity we ignore the exponential
correction term in their eq.(17).  This then gives an approximation
for the cross section for exchanging the first component, normalised
by an expression proportional to the right side of our eq.(1).
Therefore it can be interpreted as being proportional to the
probability of exchange, given that an encounter has occurred.  We
normalise this expression so that the probability of exchange of one
specific component is $1/3$
for equal masses.  The sum of the resulting probabilities of
exchanging either component may exceed unity (if $m_3$ is very large)
and in such a case we reduce both probabilities so that their sum is
unity.

If an exchange has occurred the total mass of the binary tends to
increase.  In this case we keep $a$ unchanged, and the increase in
mass hardens the binary.  If there is no exchange the binding energy
of the binary is increased in accordance with the differential cross
section given by Spitzer (1987, eq.[6-27]).  The new eccentricity is
chosen randomly from the thermal distribution.

\subsection{Stellar Evolution}

Before any encounter takes place the evolution of the binary has to be
initialised and the first single star given a unique identifier.  Whenever an encounter is
scheduled to take place, the evolution of both the binary and the
single star are updated to the time of the encounter.  If, during this
evolution step, the binary is dissociated by a supernova or if the two
stars coalesce, we stop the simulation.  Otherwise, after the
evolution step the orbital parameters of the binary ($a$ and $e$) and
the masses ($m_1$ and $m_2$) of the two stars are returned to the
dynamical part of the code.

Then the stellar evolution of the new third body is initialised. It is
evolved to the moment the encounter is scheduled and the stellar
parameters are computed.   Then we continue by
implementing the encounter, as described in the previous section. 

We consider two distinct stellar/binary evolution modules. 
The first is \SeBa\, (Portegies Zwart et al. 2001) which is part of the {\tt starlab} 
package and provides stellar and binary evolution for the {\tt kira} 
$N$-body code. 
The second is BSE (Hurley, Tout \& Pols 2002) which provides the same function 
for the {\tt NBODY4} code (Aarseth 2004). 
Both of these modules also work as stand-alone binary population synthesis 
packages. 
In each case the binary evolution is prescription based -- in order to facilitate 
the rapid computation of many binaries -- although there are differences in the 
implementation (see references and Section 3.1). 
For \SeBa\, the underlying stellar evolution is provided by the evolution formulae 
presented by Eggleton, Fitchett \& Tout (1989) while BSE uses the SSE stellar 
evolution package (Hurley, Pols \& Tout 2000).

\subsection{The Interface}

The interface between the dynamics and the stellar evolution is
separated from both program parts. Though written in C++ (\SeBa) or
Fortran (BSE) it can be linked-in to either FORTRAN, C, C++, etc.\,
code.  The interface is limited to specific operations which are
generally required for the communication. These include routines to
pass or request information, force an update and request for update
times. See the Appendix for details.

\section{Application}

\subsection{The evolution of a single binary in a dense star cluster}\label{sect:Sbinary}

McScatter is meant merely to illustrate how to construct an interface
between a dynamical code and \SeBa\, or BSE.  It is not intended for
dynamical investigations, for which a much more carefully constructed
code would be needed (cf. Ivanova et al 2004).  Nevertheless the
following exercise illustrates in a qualitative way some of the
interactions between dynamics and stellar evolution

We use the code which has been presented in this paper to model a
particular binary in the core of a large star cluster, in a similar
environment as is presented in Portegies Zwart et al.\,(1997), where
binary-binary encounters are ignored.  For initial binary parameters
we select the semi-major axis $a=4000$\,\rsun\, eccentricity $e=0.6$,
primary and secondary masses $M=2$\,\msun, and $m=1$\,\msun,
respectively.  In isolation this binary evolves in about 9.8\,Gyr
(11.2\,Gyr) to a pair of carbon-oxygen white dwarfs with masses
0.64\,\msun\, (0.64\,\msun) and  0.54\,\msun\, (0.52\,\msun) with
semi-major axis $a \simeq 10,200$\,\rsun (8600\,\rsun). Interestingly
enough the simulation with BSE (in brackets) resulted in a slightly
reduced eccentricity of 0.55, whereas \SeBa\, reported no change. This
change is mainly caused by differences in the treatment of tidal
circularization, and also largely explains the difference in the final
semi-major axes. (See fig.\,\ref{fig:SeBa} for a graphic presentation
of the binary evolution for \SeBa).  The evolution timescales also
differ and this is directly related to the distinct nature of the
evolution prescriptions.  For example, the SSE package used in BSE was
constructed from stellar models that included convective overshooting
and updated opacity tables in comparison to the Eggleton, Fitchett \&
Tout (1989) models used by \SeBa.  We note that solar metalicity is
assumed for the stars in our chosen binary.

\begin{figure}
\psfig{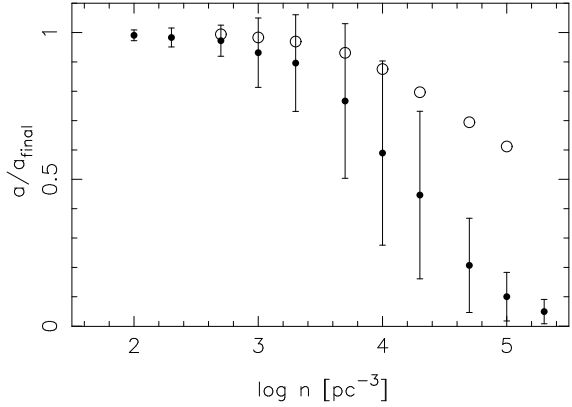} 
\caption[]{Distribution of semi major axes of the binary at an age of
12\,Gyr as a function of the stellar number density in the simulated
environment.  The semi-major axes are normalized to $a_{\rm final}$,
the value resulting from a zero-density environment. The $\bullet$
symbols indicate the average semi-major axis resulting from the
\SeBa\, implementation, and the bars indicate the dispersion in the
distribution.  The circles give the average result of the BSE
implementation.  }
\label{fig:na_std}
\end{figure}

Now we evolve this binary with McScatter with a stellar density of the
background population ranging from $n = 100$pc$^{-3}$ to $n \simeq
10^5$pc$^{-3}$, and velocity dispersion $\sigma =10$\,km\,s$^{-1}$.
For each selected density we initialize the binary $10^3$ times and
evolve it against a background of single stars, which are taken from a
Salpeter initial mass function between 0.1\,\msun\, and 100\,\msun.
In figure\,\ref{fig:na_std} we show the mean of the final orbital
separation of the binary after a 12\,Gyr evolution in a cluster as a
function of the background density.  Each simulation takes less than 2
seconds with either code for the above mentioned initial conditions
with $n = 2000$ stars/pc$^3$ on a 3.2GHz P4 pc.

The normalised final semi-major axis in the \SeBa\, implementation is on average
considerably smaller than for the BSE implementation, as is evident in
figure\,\ref{fig:na_std}. This shows clearly the cascading effect of a
larger final semi-major axis for \SeBa\, in the absence of dynamical
encounters. This gives a larger scattering cross-section (see Eq. 1) and 
tends to harden the binaries in the \SeBa\, runs more
efficiently than in the BSE simulations.  In addition,
the average black hole mass
in the BSE runs is about 8\,\msun\, with a maximum of 11\,\msun,
whereas in \SeBa\, they have a much broader distribution ranging from
about 5\,\msun\, to well over 30\,\msun. Relatively massive black holes
tend to effectively harden binaries for the entire duration of the
evolution, and this effect also causes the \SeBa\, final semi-major
axes to be considerably smaller than in the BSE runs, in particular for
the higher density environments.
 
\begin{figure}
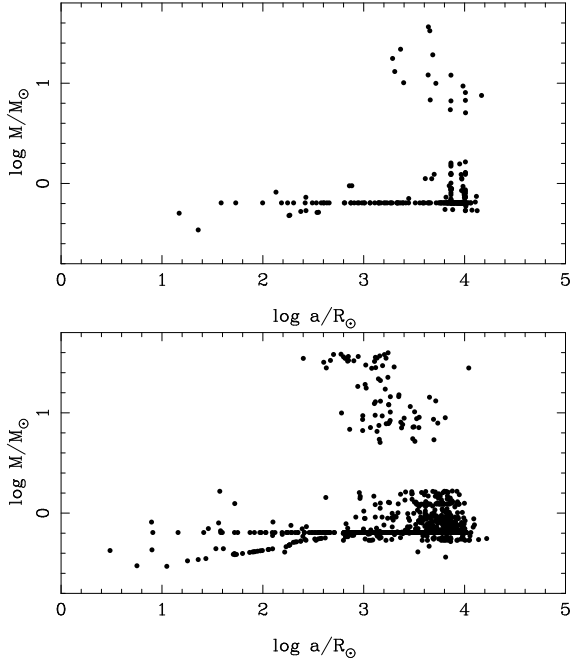

\psfig{figure=./fig_n2000_aM.ps,width=7.5cm,angle=-90} 
\psfig{figure=./fig_n20000_aM.ps,width=7.5cm,angle=-90} 
\caption[]{Distribution of semi-major axis and primary mass at an age
of 12\,Gyr for the standard binary of $a=4000$\,\rsun\,, $M=2$\,\msun\,
and $m=1\,\msun$, using SeBa.  Top panel for a cluster density of 2000pc$^{-3}$,
bottom panel for a density of 20,000pc$^{-3}$.}
\label{fig:aM}

\end{figure}

\begin{figure}
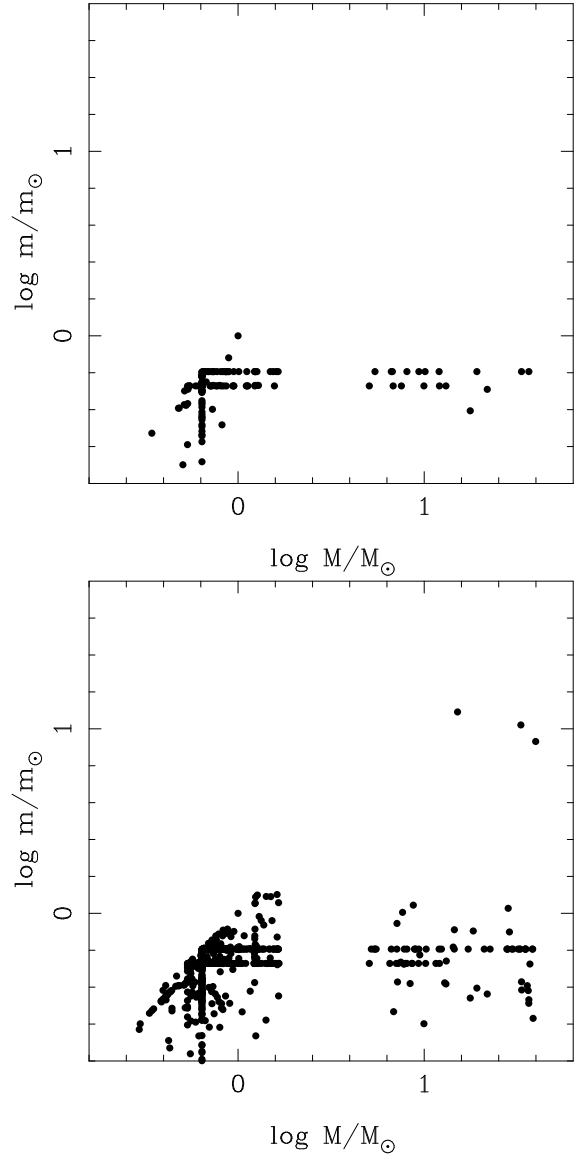

\psfig{figure=./fig_n2000_Mm.ps,width=7.5cm,angle=-90} 
\psfig{figure=./fig_n20000_Mm.ps,width=7.5cm,angle=-90} 
\caption[]{Distribution of primary and secondary masses at an age of
12\,Gyr for the standard binary, using SeBa.  Top panel for a cluster density of
2,000pc$^{-3}$, bottom panel for a density of 20,000
stars/pc$^{-3}$.  }
\label{fig:Mm}
\end{figure}

Figures \,\ref{fig:aM} and \ref{fig:Mm} show distributions of the
system parameters at an age of 12\,Gyr for environments with a
constant density of $n=2000$ pc$^{-3}$ and $n=20000$ pc$^{-3}$, using SeBa.  These
distributions can be understood quite easily.  The larger scatter in
the plot from the denser system is a direct result of the higher
encounter rate. The apparent lines of preferred solutions are somewhat
curious at first, but the trends become clear if one imagines that a
limited number of solutions are preferred after each encounter.  For
example; the concentration of primary masses around a mass of
0.64\,\msun\, ($\log M/\msun \simeq -0.19$) in both panels of Figure
\,\ref{fig:aM} are the result of encounters where the primary star is
preserved but subsequent encounters led to variations in the orbital
separation.  The clump of stars with $\log M/\msun \apgt 0.5$ are
binaries in which the primary is replaced by a black hole.

Also figure \,\ref{fig:Mm} shows interesting correlations. Here of
course, the primary and secondary stars both have a finite probability
to remain in the binary, giving rise to two horizontal as well as two
vertical correlations near $\log M/\msun = -0.19$ and $-0.27$.


\subsection{The {\tt Roche} visualization tool}

\begin{figure}
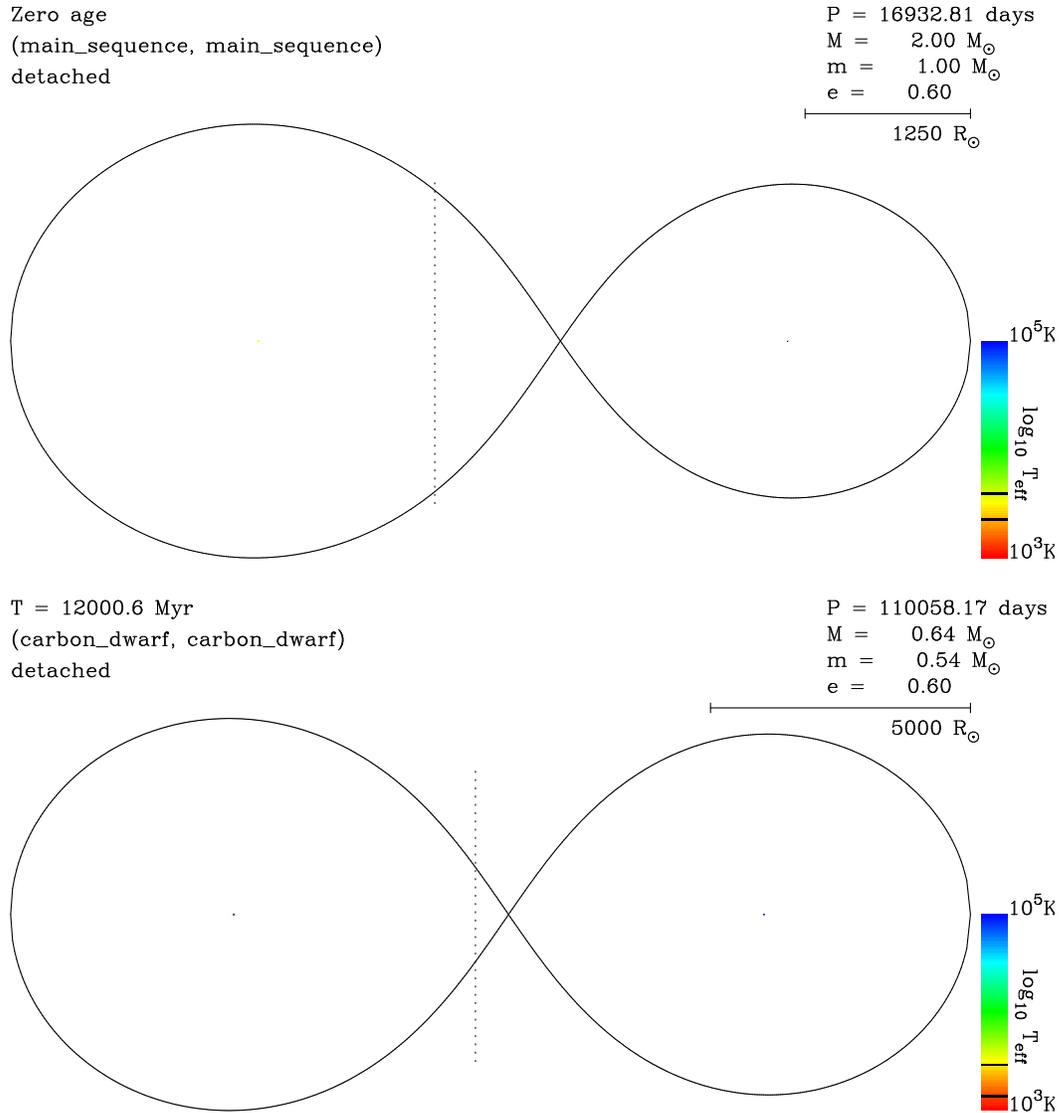

\psfig{figure=SeBa_0001.ps,width=\linewidth,angle=-90} 
\vspace*{0.5cm}
\psfig{figure=SeBa_0012.ps,width=\linewidth,angle=-90} 
\caption[]{First and and last instance of the binary as it would
evolve in isolation. Roche-lobe overflow does not occur because the
binary is far too wide.  Most of the legend is self-explanatory.  The
centre of mass lies on the vertical line. The stars in their
Roche-lobes are printed to scale, rendering the main-sequence stars
and white dwarfs hardly visible. The vertical shaded bar to the lower
right gives an indication of the effective temperature of the stars.
}
\label{fig:SeBa}
\end{figure}

The output of McScatter can be read directly by {\tt
Roche}\footnote{Roche is available via {\tt
http://www.manybody.org/manybody/roche.html}}, a visualization and
analysis tool for drawing Roche lobes of evolving binaries
\footnote{Roche requires pgplot to be installed on your system, which
is available via {\tt
http://www.astro.caltech.edu/$\sim$tjp/pgplot/}}. Roche can be used as a
stand alone program reading data from the command line or from a file
which is generated by the \SeBa\, binary evolution package and the BSE
interface for McScatter.  An example of the initial and final states
of the binary from \S\,\ref{sect:Sbinary} is illustrated in
fig.\,\ref{fig:SeBa}. Clearly, this binary has a rather unremarkable
evolution, as both stars evolve without perturbing each other
significantly.

The possibility of dynamical encounters makes the output considerably
more interesting and also more complicated.  The output of McScatter
can be read by {\tt Roche}, and an example is presented in
fig.\,\ref{fig:Roche}.

McScatter does not compute the orbits of 3-body encounters, and {\tt
Roche} is therefore not able to draw interesting spaghetti diagrams as
presented in Fig.19.2 of Heggie \& Hut (2003). Instead of this {\tt
Roche} reports strong encounters as text, as in fig.\,\ref{fig:Roche}.

\begin{figure}
\psfig{figure=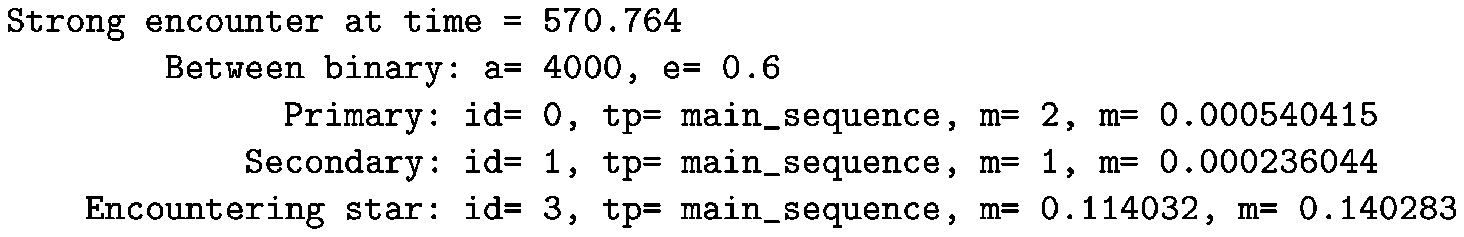,width=0.8\linewidth} 
\vspace*{0.5cm}
\psfig{figure=SeBa_0003.ps,width=1.0\linewidth,angle=-90} 
\vspace*{0.5cm}
\psfig{figure=SeBa_0005.ps,width=1.0\linewidth,angle=-90} 
\caption[]{ Evolution of the same binary is in fig.\,\ref{fig:SeBa} in
a stellar system with a density of $n = 2000$\,stars\,pc$^{-3}$. The
following sequence of events occurs. Here Roche-lobe overflow occurs
because the encounter with a low mass star at $t \sim 571$\,Myr
induces a high eccentricity on the binary. Subsequent tidal
interactions circularize the binary and the orbit shrinks. The first
phase of mass transfer occurs at $t \sim 938$\,Myr, followed by a
second phase of mass transfer at $t \sim 9.7$\,Gyr (continued on the
next page). }

\label{fig:Roche}
\end{figure}
\begin{figure}
\psfig{figure=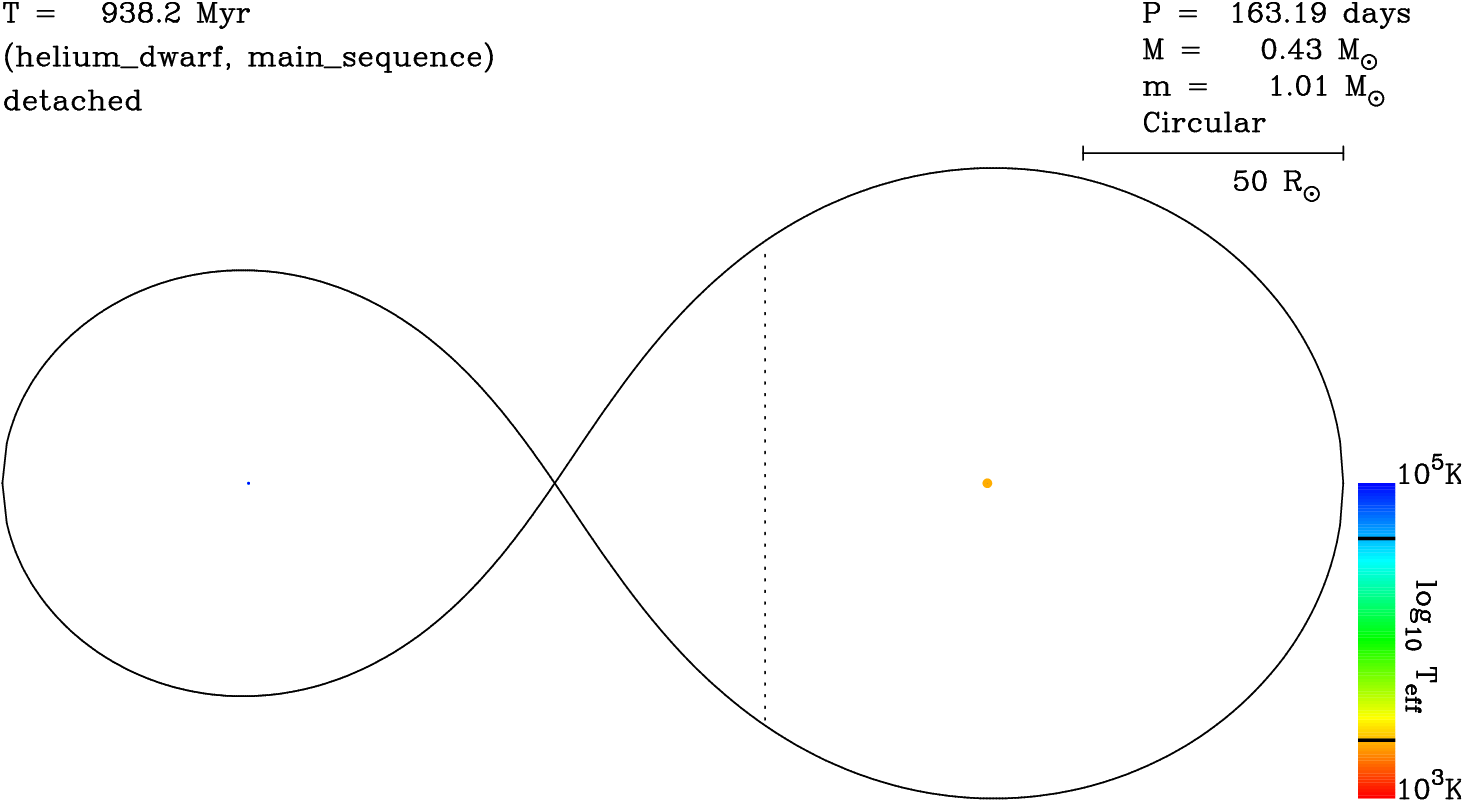,width=1.0\linewidth,angle=-90} 
\vspace*{0.5cm}
\psfig{figure=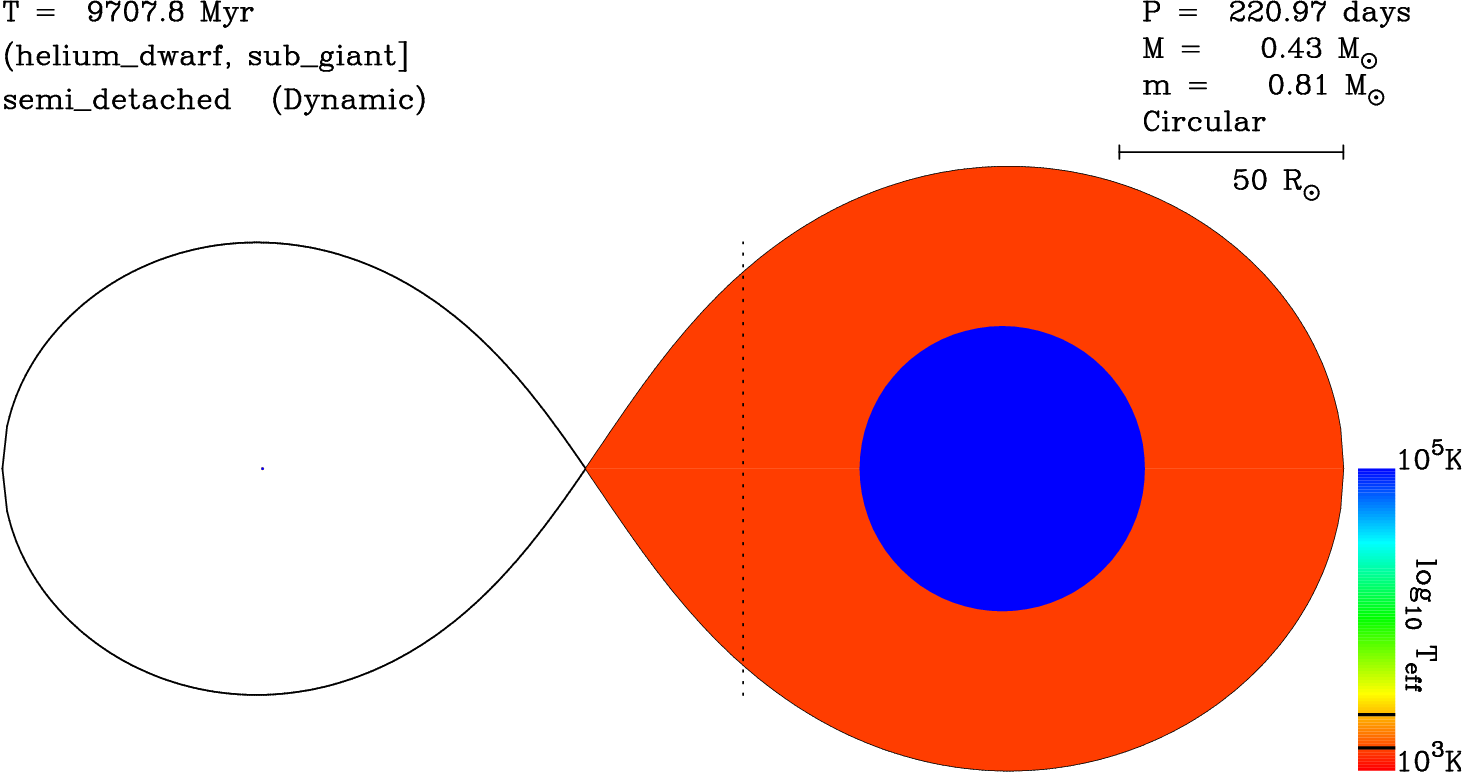,width=1.0\linewidth,angle=-90} 
\vspace*{0.5cm}
\psfig{figure=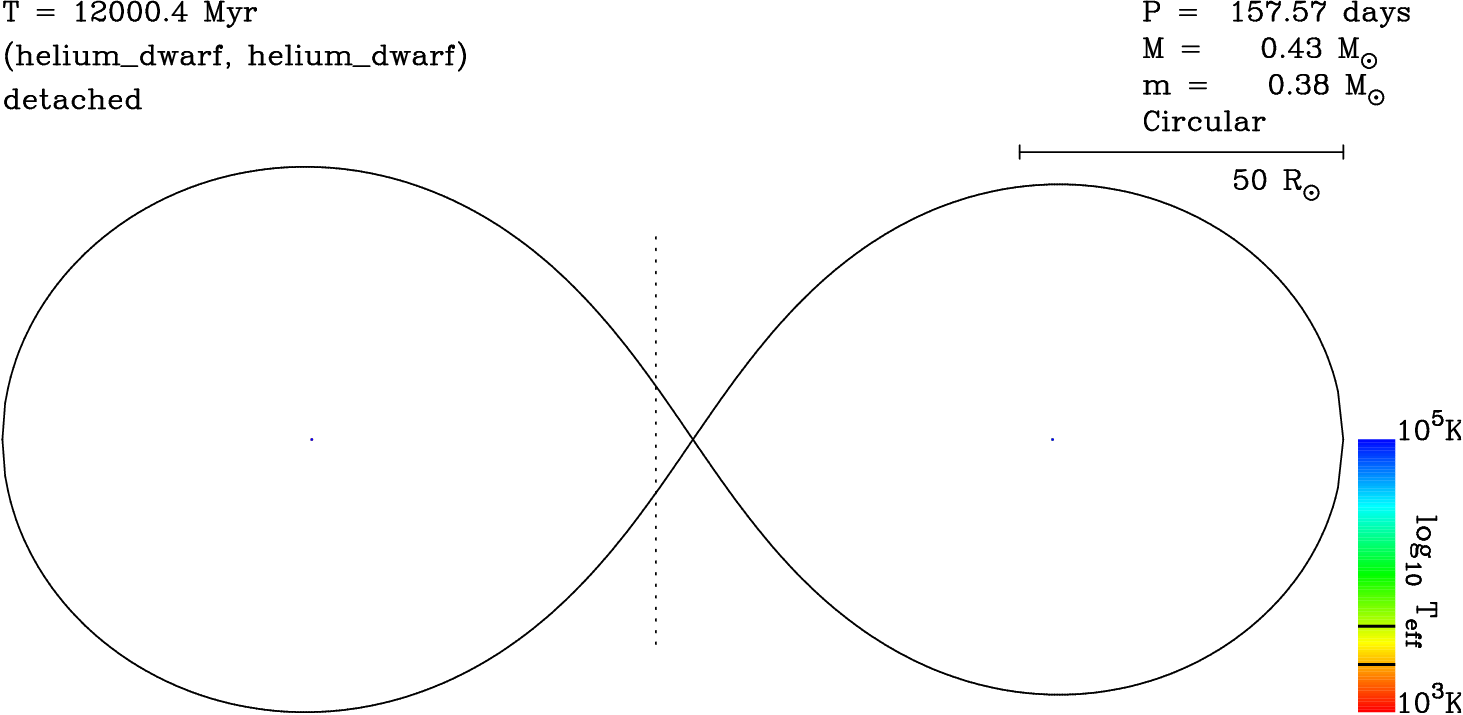,width=1.0\linewidth,angle=-90} 
\label{fig:Roche}
\end{figure}

\section{Conclusions}

This paper presents an interface between stellar dynamics and stellar
evolution which is intended for use in the context of dense stellar
systems.  For purposes of illustration, however, we have presented it
in a situation where the dynamics is as simple as possible, i.e. a
simplified model for three-body scattering between a binary and a
single star.  In particular, this illustrates how the interface
operates even in situations where exchanges take place.  The resulting
toy code is called {\tt McScatter}, for Monte Carlo Scattering.

To illustrate the power of our interface, we have implemented it for
two independent models for the internal evolution of binary and single
stars.  These are {\tt SeBa}, the stellar evolution package within
starlab (see Portegies Zwart et al 2001), and {\tt BSE} (Hurley, Tout
\& Pols 2002), which is usually used in conjunction with Aarseth's
N-body codes (Aarseth, 2004).  These codes as they stand are not able
to model such rich systems as massive globular clusters or galactic
nuclei, but it is intended that our interface will be employed in
other stellar dynamics codes (such as Monte Carlo codes) which can.

To highlight the differences which can arise, we have considered the
scattering of a binary within a field of single stars.  We have shown
that the different treatment of tidal effects in the two stellar
evolution packages can lead to substantial differences in the
dynamical evolution of a binary.  While our application is highly
simplified, there is no doubt that comparable differences would arise
in realistic situations.

\section*{Appendix}

Here we list the basic interface routines for stellar evolution and
stellar dynamics.

\section*{A ~~~ {\tt SeBa}}

As in {\tt starlab} the stellar dynamics is
supposedly the driver for the slaved stellar evolution. To assure
two-way communication there is a separate set of routines to force the
dynamics to update the stellar evolution, or to inform the dynamics of
the state of a star or binary.

There are five families of routines, initialization, {\tt get}- and
{\tt set}- operators, the core evolution routines, I/O routines and a
miscellaneous group. The routine names start with a few characters
which identify the class to which the routine belongs.  These classes
are: {\tt init\_}, {\tt get\_}, {\tt set\_}, {\tt ev\_} and {\tt
put\_}.

\vskip0.5truein

{\sl List of routines}

\begin{itemize}
  \item[$\bullet$]{\tt init}: Initialization routines \\
    {\tt void init\_stars(int {\rm *n}, int {\rm id[]}, real {\rm mass[]})}\\
    {\tt void init\_binaries(int {\rm *n}, int {\rm id[]}, 
         real {\rm sma[]}, real {\rm ecc[]}, 
	 real {\rm mprim[]}, real {\rm msec[]})} \\
	 initialization of a list of {\tt n} stars or binaries
	 identified with arrays of length $n$ for the identity {\tt
	 id} and zero age mass {\tt mass}, or in the case of a binary
	 with primary mass {\tt mprim}, secondary mass {\tt msec}, the
	 semi-major axis {\tt sma} and the eccentricity {\tt ecc}.

  \item[$\bullet$]{\tt get/set}: operators to initialize and inquire the value
        which is identified in the second part of the function
        name. These routines are called for each individual star or
        binary separately. \\
    {\tt void get\_mass(int {\rm *id}, real {\rm *mass})} \\
    function reads the mass from the stellar evolution package to be
    used in the dynamics code. \\
    {\tt void set\_sma(int {\rm *id}, real {\rm *sma})} \\
    function initializes the orbital separation of a binary after it
    was affected by a strong encounter. \\
    Other available functions are: 
    {\tt set/get\_sma(...)},
    {\tt set/get\_ecc(...)},
    {\tt get\_radius(...)},
    {\tt get\_ss\_type(...)},
    {\tt get\_bs\_type(...)},
    {\tt get\_loid(...)},
    {\tt get\_hiid(...)}, 
    {\tt get\_loid\_mass(...)},
    {\tt get\_hiid\_mass(...)}, 
    {\tt get\_ss\_updatetime(...)},
    {\tt get\_bsi\_updatetime(...)}. \\
    For some parameters {\tt set\_} operations are not externally available.
    Note that {\tt get\_loid/\_hiid} are available to prevent
    confusion of the identity of the {\em primary} and {\em secondary}
    stars (which may or may not be the stars of lower or higher {\tt id}).

  \item[$\bullet$]{\tt ev}: routines to force evolution \\
    {\tt void ev\_stars(int {\rm *n}, int {\rm *id}, real {\rm *time})} \\
    {\tt void ev\_binaries(int {\rm *n}, int {\rm *id}, real {\rm *time})} \\ 
    Forces a list of $n$ {\tt \_stars/\_binaries} to evolve to time
    {\rm *time}.  \\

  \item[$\bullet$]{\tt put}: input/output routines. \\
    The current implementation contains only output routines: \\
    {\tt void out\_star(int {\rm *id})}, \\
    {\tt void out\_binary(int {\rm *id})} and \\
    {\tt void out\_scatter(int {\rm *idb}, int {\rm *idt}, real {\rm *time},
		      bool {\rm *write\_text})}\footnote{The output of
    this routine can be read directly by {\tt Roche}.}

  \item[$\bullet$]other] non-classified routines. \\
    {\tt void binary\_exists(int {\rm *id}, int {\rm *exit})} \\
    {\tt void morph\_binary(int {\rm *idb}, int {\rm *idt}, 
      real {\rm *a\_factor}, real {\rm *ecc}, int {\rm *outcome})}\\
function to reinitialise the binary after an exchange.

\end{itemize}

Some routines are available in plural form by adding a {\tt s} to the
end of the function name, these operate on arrays of stars of binaries
rather than on an individual object.

\section*{B. ~~~~ BSE}

The routines for single- and binary-star evolution are contained in a
precompiled library of Fortran subroutines and functions: libstr.a .
At present versions are available for OSX, Linux PC and alpha.  These
are accessed through an interface, which replaces the \SeBa\, interface
described in the previous subsection.

\vskip 0.25truein
{\sl List of subroutines}

\begin{itemize}
\item[$\bullet$]{\tt evStar, evBinary}: 
  These  subroutines  provide the link between the 
      main program and the SSE/BSE library (or module). These routines 
      evolve the star (or binary) forward by a user specified interval. 
      The library routines return a recommended update timestep based on 
      the evolution stage of the star, and the parameters dmmax (maximum 
      allowed change in mass) and drmax (maximum allowed change in radius). 
      The main program may or may not utilise this timestep. 
      (This routine also produces some of the output which can be read 
	directly by {\tt Roche}.)
\item[$\bullet$]{\tt initPar}:  Input options for the SSE/BSE library are set in
	this subroutine, and these may be modified by the user (see
	the routine for an explanation of these options).  The common
	blocks that convey these options to the library are declared
	in const\_bse.h. 
\item[$\bullet$]{\tt initStar, initBinary}: These are used to initialise single
      stars and binaries (respectively).  {\tt  initBinary} in turn
      initializes two single stars.
\item[$\bullet$]{\tt         ssupdatetime}:  Provides next update time for star i 
\item[$\bullet$]{\tt         bsupdatetime}:  Calls ssupdatetime using index of primary star 
\item[$\bullet$]{\tt         binaryexists}:  Determines if binary remains bound
  based on the index of binary type
\item[$\bullet$]{\tt         getLabel}:     Text label associated with index of stellar type 
\item[$\bullet$]{\tt         getLabelb}:    Text label associated with index of binary type  
\item[$\bullet$]{\tt         printstar}:    Formatted output for a single star 
\item[$\bullet$]{\tt         printbinary}:  Formatted output for a binary 
\item[$\bullet$]{\tt         print\_roche\_data}:  Primary output for the 
	{\tt Roche} package

\end{itemize}

The quantities and units used in BSE are explained in the preamble of the
file interface\_bse.f. In summary there are, as in \SeBa,
routines for setting and fetching such variable as the mass and radii
of the stars and other parameters for the binary.  The arrays that
store these quantities are declared in interface\_bse.h (where the
user may choose to alter the size of these arrays, i.e. nmax).  An
additional file const\_bse.h contains parameters needed by the BSE
subroutine library.

\section*{Acknowledgments}
DCH warmly thanks the University of Amsterdam for its hospitality on
several occasions.  This work was made possible by financial support
from the Nederlandse Onderzoekschool voor Astronomie (NOVA) and the
Royal Netherlands Academy of Arts and Science (KNAW).  We thank the
referee for suggesting a number of improvements to the paper.

\section*{References}

{

\leftskip=0.2truein 
\parindent=-0.2truein

\hskip-0.2truein 

Aarseth S.J., 2004.  {\sl Gravitational $N$-Body Simulations.  Tools
  and Algorithms} (CUP, Cambridge)
  
Eggleton P.P.,  Fitchett M., Tout C.A., 1989, ApJ, 347, 998

Heggie D.C., Hut P., 2003.  The Gravitational Million-Body Problem
(CUP, Cambridge)

Heggie D.C., Hut P., McMillan S.L., 1996, ApJ, 467, 359

Hurley J. R., Pols O.R., Tout C. A., 2000, MNRAS, 315, 543

Hurley J. R., Tout C. A., Pols,O.R., 2002, MNRAS, 329, 897

Ivanova N.S., Belczynski K., Fregeau J.M., Rasio F.A., 2004, MNRAS, 358, 572

Portegies Zwart S.F., Hut, P., McMillan S.L.W., Verbunt, F., 1997,
A\&A, 328, 143

Portegies Zwart S.F., McMillan S.L.W., Hut P., Makino J., 2001, MNRAS,
321, 199

Spitzer L., Jr., 1987. Dynamical Evolution of Globular Clusters (PUP,
Princeton) 

}

\end{document}